\documentclass[twocolumn]{svjour3}

\usepackage{graphics}
\usepackage{amssymb}

\begin{document}

\title{Magnetic response of optimally doped Pr$_{1-x}$LaCe$_x$CuO$_4$}

\author{A. Sherman}

\institute{Institute of Physics, University of Tartu, Riia 142,
51014 Tartu, Estonia\\
              Tel.: +372-7-374616\\
              Fax: +372-7-383033\\
              \email{alexei@fi.tartu.ee}
}

\date{Received: date / Accepted: date}

\maketitle

\begin{abstract}
The magnetic susceptibility of the optimally doped Pr$_{1-x}$LaCe$_x$CuO$_4$ in the superconducting state is calculated using the $t$-$J$ model of Cu-O planes, Mori's projection operator technique and the dispersion of electron bands derived from photoemission experiments. The electron band folding across the antiferromagnetic Brillouin zone border, which is inherent in the crystal, leads to a commensurate low-frequency response. The same band folding causes the appearance of a supplementary spin-excitation branch. The coexistence of the two spin-excitation branches explains two maxima observed in the frequency dependence of the susceptibility. The calculated momentum and frequency dependencies are close to experimental observations. Similarities and differences in the magnetic responses of electron- and hole-doped cuprates are discussed.
\end{abstract}

\section{Introduction}
Magnetic responses of $n$- and $p$-type cuprate perovskites are essentially different. In the former crystals, the low-frequency susceptibility is commensurate, while in the latter it is incommensurate, heaving peaks at momenta, which differ from the antiferromagnetic (AF) wave vector ${\bf Q}=(\pi,\pi)$ \cite{Armitage,Fujita11} [hereafter, I use the two-dimensional (2D) notations for wave vectors in a Cu-O plane with the lattice spacing set as the unit of length]. In $n$-type cuprates the dispersion of the susceptibility maxima resembles a cone with the apex point at the frequency $\omega=0$ and at the momentum {\bf Q} \cite{Wilson,Fujita06}. In $p$-type cuprates this dispersion has the hourglass shape with the waist at {\bf Q} and at the frequency $25-50$~meV in moderately doped crystals \cite{Fujita11}.

The susceptibility of $n$-type cuprates was calculated in the random phase approximation in Refs.~\cite{Li,Ismer}. The obtained result depends largely on the momentum dependence of the electron four-point vertex $U_{\bf q}$, which choice is, to a certain extent, arbitrary in this approach. It is known that $U_{\bf q}$ peaked at {\bf Q} tends to suppress incommensurability. Such a vertex was chosen in Refs.~\cite{Li,Ismer}. However, even with this vertex and with simple tight-binding approximations used for the electron dispersion in these works the low-frequency susceptibility demonstrates weak incommensurability \cite{Li}. In the consideration of hole-doped cuprates the vertex is usually set to a constant (see, e.g., \cite{Norman} and references therein). The application of such a vertex to $n$-type cuprates leads to pronounced incommensurability, which is inconsistent with experimental observations \cite{Kruger}.

In this article, the $t$-$J$ model of Cu-O planes is used to calculate the dynamic magnetic susceptibility of Pr$_{1-x}$LaCe$_x$CuO$_4$ (PLCCO) in the superconducting state. For $x=0.11-0.12$ the magnetic response of this crystal was studied in a series of neutron scattering experiments \cite{Armitage,Fujita11,Wilson,Fujita06,Fujita08,Zhao,Zhao11}, the electron dispersion was derived \cite{Das} from photoemission data \cite{Matsui}, and estimates for the value of the superconducting gap were obtained from high-resolution scanning tunneling microscopy measurements \cite{Zhao11,Niestemski}. The use of the $t$-$J$ model allows one to take proper account of strong electron correlations inherent in $n$-types cuprates. To calculate Green's functions constructed from Hubbard operators of the model Mori's projection operator technique \cite{Mori} is used. For $p$-type cuprates this approach allowed us to reproduce the observed momentum and frequency dependencies of the susceptibility using hole dispersions derived from photoemission  \cite{Sherman12,Schreiber}. The cases of superconducting and pseudogap phases were considered.

The electronic structure of $n$-type cuprates is characterized by band folding across the AF Brillouin zone border \cite{Armitage,Das}. This fact plays a central role in the formation of the commensurate low-frequency response. The momentum dependence of the low-frequency susceptibility is governed by the spin-excitation damping, which peaks at {\bf Q} due to the band folding. This peculiarity of the electron-doped cuprates leads also to the appearance of a supplementary spin-excitation branch. As a consequence the dispersion of the susceptibility maxima has the shape of a cone. Its upper part is formed by the nested into each other branches of the usual and supplementary spin excitations. The part near the apex point is determined by the spin-excitation damping. The same two parts can be singled out in the dispersion of hole-doped cuprates, with the difference that the supplementary spin excitations are lacking there and the low-frequency spin-excitation damping peaks at incommensurate momenta. In the frequency dependence of the susceptibility, the usual and supplementary spin excitations manifest themselves as two maxima or a maximum and a shoulder. Such frequency dependencies were recently observed in PLCCO \cite{Zhao,Zhao11}.

\section{Main formulas}
Formulas for the magnetic susceptibility of the 2D $t$-$J$ model were derived in Ref.~\cite{Sherman12} using Mori's projection operator technique. In this approach, the formally exact expression for the susceptibility can be obtained \cite{Mori}
\begin{equation}\label{chi}
\chi({\bf k}\omega)=-\frac{h_{\bf k}}{\omega^2-\omega\Pi({\bf k}\omega)-\omega_{\bf k}^2}.
\end{equation}
Here {\bf k} is the 2D wave vector, other parameters are expressed through correlators of the spin-$\frac{1}{2}$ operators $s^z_{\bf k}$, $s^\pm_{\bf k}$ and their time derivatives. In the case of the $t$-$J$ model these parameters read
\begin{eqnarray}
h_{\bf k}&=&4\left(\tilde{t}_0F_1+J|C_1|\right)\left(1-\gamma_{\bf k}\right), \label{hk}\\[1ex]
\omega^2_{\bf k}&=&16J^2\alpha|C_1|\left(1+\frac{\tilde{t}_0F_1}{J\alpha|C_1|} \right)\left(1-\gamma_{\bf k}\right)
\left(\delta+1+\gamma_{\bf k}\right), \nonumber\\[-1.5ex]
&&\label{wk}
\end{eqnarray}
where $\tilde{t}_0$ and $J$ are the hopping and exchange constants between neighboring sites in the $t$-$J$ Hamiltonian, $\gamma_{\bf k}=\left[ \cos(k_x)+\cos(k_y)\right]/2$, $$F_1=\frac{1}{N}\sum_{\bf k}\gamma_{\bf k}\left\langle a^\dagger_{\bf k\sigma}a_{\bf k\sigma}\right\rangle\;{\rm and}\; C_1=\frac{1}{N}\sum_{\bf k}\gamma_{\bf k}\left\langle s^{+}_{\bf k}s^{-}_{\bf k}\right\rangle$$
are correlators of the electron $a^{(\dagger)}_{\bf k\sigma}$ and spin operators on the neighboring sites, $\sigma$ is the spin projection, the angular brackets denote the statistical averaging and $N$ is the number of sites. The parameter $\alpha$ serves for correcting the decoupling procedure \cite{Kondo,Shimahara} used for deriving Eqs.~(\ref{hk}) and (\ref{wk}). For small electron concentrations $\alpha\approx 1.7$ \cite{Sherman03}. The quantities $\omega_{\bf k}$ and $\Pi({\bf k}\omega)$ are the frequency of spin excitations and their polarization operator, which contains contributions from interactions with electrons and with other spin excitations.
The parameter $\delta$ in Eq.~(\ref{wk}) describes a gap in the spin-excitation spectrum near {\bf Q} due to temperature fluctuations \cite{Shimahara} and/or the interaction with electrons \cite{Sherman03}. The magnitude of this gap is directly connected with the correlation length of the short-range AF order, and it grows with the electron concentration $\bar{x}\approx x$ and temperature $T$. In the case of hole doping the magnitude of the gap determines the frequency $\omega_r$ of the waist in the hourglass dispersion of the susceptibility maxima \cite{Sherman12}.

Calculations are essentially simplified for zero temperature. For this case, the polarization operator reads
\begin{eqnarray}
&&\Pi({\bf k}\omega)=\Pi_1({\bf k}\omega)+\Pi_2({\bf k}\omega)+\Pi_3({\bf k}\omega),\label{pi}\\
&&\Pi_1({\bf k}\omega)=\frac{J|C_1|Z^2}{4N^2h_{\bf k}}\sum_{\bf qq'}\sum_{\tau\tau'}\left[2f_1^2({\bf kqq'})+f_3^2({\bf kqq'})\right]\nonumber\\
&&\quad\times\frac{(1-\gamma_{\bf k'})({\cal E}_{\bf q}+\tau\varepsilon^-_{\bf q})({\cal E}_{\bf q'}+\tau'\varepsilon^-_{\bf q'})}{\omega_{\bf k'}{\cal E}_{\bf q}{\cal E}_{\bf q'}E_{\bf q\tau}E_{\bf q'\tau'}(E_{\bf q\tau}+E_{\bf q'\tau'}+\omega_{\bf k'})}\nonumber\\
&&\quad\times\Big[(E_{\bf q\tau}+\varepsilon^+_{\bf q}+\tau{\cal E}_{\bf q})(E_{\bf q'\tau'}-\varepsilon^+_{\bf q'}-\tau'{\cal E}_{\bf q'})\nonumber\\
&&\quad+\Delta_{\bf q}\Delta_{\bf q'}\Big]\left(\frac{1}{\omega+E_{\bf q\tau}+E_{\bf q'\tau'}+\omega_{\bf k'}+i\eta}\right.\nonumber\\
&&\quad+\left.\frac{1}{\omega-E_{\bf q\tau}-E_{\bf q'\tau'}-\omega_{\bf k'}+i\eta}\right),\label{pi1}\\
&&\Pi_2({\bf k}\omega)=\frac{32J^3|C_1|^3}{N^2h_{\bf k}}\sum_{\bf qq'}f_4^2({\bf kqq'})\nonumber\\
&&\quad\times\frac{(1-\gamma_{\bf k'})(1-\gamma_{\bf q})(1-\gamma_{\bf q'})}{\omega_{\bf k'}\omega_{\bf q}\omega_{\bf q'}(\omega_{\bf k'}+\omega_{\bf q}+\omega_{\bf q'})}\nonumber\\
&&\quad\times\left(\frac{1}{\omega+\omega_{\bf q}+\omega_{\bf q'}+\omega_{\bf k'}+i\eta}\right.\nonumber\\
&&\quad+\left.\frac{1}{\omega-\omega_{\bf q}-\omega_{\bf q'}-\omega_{\bf k'}+i\eta}\right),\label{pi2}\\
&&\Pi_3({\bf k}\omega)=\frac{Z^2}{8Nh_{\bf k}}\sum_{\bf q}\sum_{\tau\tau'}f_2^2({\bf kq})\nonumber\\
&&\quad\times\frac{({\cal E}_{\bf k+q}+\tau\varepsilon^-_{\bf k+q})({\cal E}_{\bf q}+\tau'\varepsilon^-_{\bf q})}{{\cal E}_{\bf k+q}{\cal E}_{\bf q}E_{\bf k+q,\tau}E_{\bf q\tau'}(E_{\bf k+q,\tau}+E_{\bf q\tau'})}\nonumber\\
&&\quad\times\Big[(E_{\bf k+q,\tau}+\varepsilon^+_{\bf k+q}+\tau{\cal E}_{\bf k+ q})(E_{\bf q\tau'}-\varepsilon^+_{\bf q}-\tau'{\cal E}_{\bf q})\nonumber\\
&&\quad-\Delta_{\bf k+ q}\Delta_{\bf q}\Big]\left(\frac{1}{\omega+E_{\bf k+q,\tau}+E_{\bf q\tau'}+i\eta}\right.\nonumber\\
&&\quad+\left.\frac{1}{\omega-E_{\bf k+q,\tau}-E_{\bf q\tau'}+i\eta}\right) \label{pi3},
\end{eqnarray}
where ${\bf k'=k-q+q'}$, $\eta\rightarrow +0$, $\tau$ and $\tau'=\pm 1$,
\begin{eqnarray*}
&&f_1({\bf  kqq'})=\frac{1}{2}\bigg[\varphi_1({\bf  q-k,k'})-\varphi_1({\bf  -q'-k,k'})\\
&&\quad+\frac{1}{2}\varphi_2({\bf -q'-k,q})+\varphi_2({\bf  -q',q})-\frac{3}{2}\varphi_2({\bf  -q',q-k})\\
&&\quad +\frac{3}{2}\varphi_2({\bf  -q,q'+k})-\frac{1}{2}\varphi_2({\bf  k-q,q'})
-\varphi_2({\bf  -q,q'})\bigg],\\
&&f_2({\bf  kq})=\frac{1+\bar{x}}{2}\bigg[\varphi_1({\bf  -q-k,k})-\frac{1}{2}\varphi_1({\bf  -q-k},{\bf 0})\\
&&\quad-\frac{1}{2}\varphi_1({\bf  q},{\bf 0})\bigg],\\
&&f_3({\bf  kqq'})=\varphi_1({\bf  -q,q-q'})-\frac{1}{2}\varphi_1({\bf  -q'-k,k'})\\
&&\quad-\frac{1}{2}\varphi_1({\bf  q-k,k'})+\frac{1}{4}\varphi_2({\bf  k-q,q'})\\
&&\quad+\frac{1}{4}\varphi_2({\bf  -k-q',q})-\frac{1}{4}\varphi_2({\bf  -q',q-k})\\
&&\quad-\frac{1}{4}\varphi_2({\bf  -q,q'+k}),\\
&&f_4({\bf  kqq'})=-\frac{1}{2}\varphi_3({\bf  -q,k+q'})-\frac{1}{2}\varphi_3({\bf  -k-q',k'})\\
&&\quad-\frac{1}{2}\varphi_3({\bf  q',k-q})-\frac{1}{2}\varphi_3({\bf  q-k,k'})+\varphi_3({\bf  q,q'-q})\\
&&\quad+\frac{1}{2}\varphi_3({\bf  -k-q',q})+\frac{1}{2}\varphi_3({\bf  q-k,-q'}),
\end{eqnarray*}
\begin{eqnarray}
\varphi_1({\bf  qq'})&=&\sum_{\bf  p}\left(\delta_{\bf  pq}-\frac{1}{N}\right)t_{\bf  p}t_{\bf  p+q'},\nonumber\\
\varphi_2({\bf  qq'})&=&\sum_{\bf  p}\left(\delta_{\bf  pq}-\frac{1}{N}\right)t_{\bf  p}J_{\bf  p+q'},\label{phi}\\
\varphi_3({\bf  qq'})&=&\sum_{\bf  p}\left(\delta_{\bf  pq}-\frac{1}{N}\right)J_{\bf  p}J_{\bf  p+q'},\nonumber
\end{eqnarray}
$t_{\bf q}$ and $J_{\bf q}$ are Fourier transforms of the hopping and exchange constants in the $t$-$J$ Hamiltonian.

Equations~(\ref{pi})-(\ref{pi3}) were derived using the following normal $A_{11}({\bf k}\omega)=-\pi^{-1}{\rm Im}\langle\!\langle a_{\bf k\sigma}|a^\dagger_{\bf k\sigma}\rangle\!\rangle$ and anomalous $A_{12}({\bf k}\omega)-\pi^{-1}{\rm Im}\langle\!\langle a_{\bf k\uparrow}|a_{\bf k\downarrow}\rangle\!\rangle$ electron spectral functions:
\begin{eqnarray}
A_{11}({\bf k}\omega)&=&\frac{Z}{4{\cal E}_{\bf k}}\sum_{\tau\tau'}\frac{{\cal E}_{\bf k}+\tau\varepsilon^-_{\bf k}}{E_{\bf k\tau}}\left[E_{\bf k\tau}+\tau'\left(\varepsilon^+_{\bf k}+\tau{\cal E}_{\bf k}\right)\right]\nonumber\\
&&\times\delta(\omega-\tau'E_{\bf k\tau}),\label{sf}\\
A_{12}({\bf k}\omega)&=&\frac{Z\Delta_{\bf k}}{4{\cal E}_{\bf k}}\sum_{\tau\tau'}\frac{{\cal E}_{\bf k}+\tau\varepsilon^-_{\bf k}}{E_{\bf k\tau}}\tau'\delta(\omega-\tau'E_{\bf k\tau}),\nonumber
\end{eqnarray}
where
\begin{eqnarray}
&&\varepsilon^\pm_{\bf k}=\frac{1}{2}\left(\varepsilon_{\bf k}\pm\varepsilon_{\bf k-Q}\right), \quad {\cal E}_{\bf k}=\sqrt{(\varepsilon^-_{\bf k})^2+\Delta^2_f},\nonumber\\[-1ex]
&&\label{bands}\\[-1ex]
&&E_{\bf k\tau}=\sqrt{\left(\varepsilon^+_{\bf k}+\tau{\cal E}_{\bf k}\right)^2+\Delta^2_{\bf k}}.\nonumber
\end{eqnarray}
The dispersions $\pm E_{\bf k\tau}$, Eq.~(\ref{bands}), correspond to  electron bands of a crystal with the superconducting gap function $\Delta_{\bf k}$ and with the band folding across the AF Brillouin zone border, which is characterized by the potential $\Delta_f$ \cite{Das,Valenzuela}. This potential and parameters of the initial tight-binding dispersion
\begin{eqnarray}\label{tbd}
\varepsilon_{\bf  k}&=&-2t_0\big[\cos(k_x)+\cos(k_y)\big]-4t_1\cos(k_x)\cos(k_y) \nonumber\\
&-&2t_2\big[\cos(2k_x)+\cos(2k_y)\big]-4t_3\big[\cos(2k_x)\cos(k_y)\nonumber\\
&+&\cos(k_x)\cos(2k_y)\big]
-4t_4\cos(2k_x)\cos(2k_y)-\mu
\end{eqnarray}
were obtained \cite{Das} by fitting the photoemission data \cite{Matsui} in  Pr$_{0.89}$LaCe$_{0.11}$CuO$_4$: $t_0=0.12$~eV,  $t_1=-0.06$~eV, $t_2=0.034$~eV, $t_3=0.007$~eV, $t_4=0.02$~eV, $\mu=-0.082$~eV and
$\Delta_f=0.14$~eV. In the below calculations, it was expected that only the nearest-neighbor exchange constant is nonzero, and its value $J=0.12$~eV \cite{Armitage}.

Equations~(\ref{pi1})-(\ref{pi3}) describe four processes of the spin excitation transformation -- its conversion to an electron-hole pair with assistance of another spin excitation [Eq. (\ref{pi1})], the decay into three spin excitations [Eq.~(\ref{pi2})] and the conversion into an electron-hole pair, direct and assisted by an electron [both processes are combined into Eq.~(\ref{pi3})]. In the considered range of parameters this latter term makes the main contribution to $\Pi({\bf k}\omega)$.

Notice that $t_0$ does not coincide with the parameter $\tilde{t}_0$ of the $t$-$J$ Hamiltonian in Eqs.~(\ref{hk}) and (\ref{wk}), since the interaction between electrons and spin excitations leads to a considerable reduction of the electron bandwidth. Based on the exact diagonalization results \cite{Tohyama} I set $\tilde{t}_0/t_0=2$ and supposed the same relation between other unrenormalized and renormalized hopping constants in Eq.~(\ref{phi}). Only the nearest-neighbor hoping constant was left in Eqs.~(\ref{hk}) and (\ref{wk}), since other kinetic terms introduce merely small corrections. Also a small third harmonic in the superconducting gap function \cite{Armitage} influences only slightly the susceptibility and the major part of calculations was carried out with the $d$-wave gap function $\Delta_{\bf  k}=\Delta[\cos(k_x) -\cos(k_y)]/2$. The gap parameter $\Delta$ varied in the range $4.8-9.6$~meV. These values are close to the estimates obtained by high-resolution scanning tunneling microscopy \cite{Zhao11,Niestemski}.

In Eq.~(\ref{sf}), the parameter $Z$ takes into account the difference of the operators $a^\dagger_{\bf k\sigma}$ and $a_{\bf k\sigma}$, which are the Hubbard operators, from the fermion creation and annihilation operators and the fact that in the case of strong electron correlations a sizable part of the spectral weight is concentrated in an incoherent part of the electron spectral function. The spectral weights of the coherent $Z$ and incoherent $Z'$ parts satisfy the relation
$$Z+Z'=\frac{1+\bar{x}}{2}.$$
Since the incoherent continuum does not produce peaks in the susceptibility, only the coherent parts of the spectral functions were taken into account in (\ref{sf}). Supposing that in the $n$-type case the ratio $Z/Z'$ is approximately the same as in the $p$-type case, $Z$ was set to $1/6$ for $\tilde{t}_0=2J$. In the calculations, the infinitesimally small quantity $\eta$ in Eqs.~(\ref{pi1})-(\ref{pi3}) was substituted by the artificial broadening $\Gamma=0.3-7$~meV. Such values are usually used in calculations of the susceptibility. The parameter $\Gamma$ influences mainly widths of maxima in the frequency dependence of the imaginary part of the susceptibility $\chi''({\bf k}\omega)$.

As mentioned above, the parameter $\delta$ defines the location of the bottom of the spin-excitation dispersion $\omega_r$, which corresponds usually to the maximum in the frequency dependence of $\chi''({\bf Q}\omega)$. The value of $\delta$ was fitted to adjust the calculated maximum to the experimental one. Obtained values of $\delta$ varied in the range $2\times 10^{-4}-4\times 10^{-3}$, which corresponds to $\omega_r$ of the order of several meV. These $\delta$ and $\omega_r$ are much smaller than the respective parameters in moderately doped $p$-type cuprates, which is the consequence of the fact that the considered crystal is close to the boundary of the long-range AF order. In this connection, it is worth noting that in some works a weak static AF order was observed in the optimally doped PLCCO at low temperatures (see, e.g., Refs.~\cite{Wilson,Fujita06}). At the same time it was pointed out that low-frequency maxima are broader than the instrumental resolution, and a finite correlation length of the AF order was obtained \cite{Wilson}, which implies a nonzero value of $\delta$. The question about possible coexistence of superconductivity and long-range AF order is still open due to the influence of sample inhomogeneity \cite{Armitage}. Following results of Refs.~\cite{Fujita08,Motoyama} in this work it is supposed that the long-range AF order is lacking in the superconducting phase and $\delta$ is finite.

\section{Results and discussion}
\begin{figure}
\centerline{\resizebox{0.95\columnwidth}{!}{\includegraphics{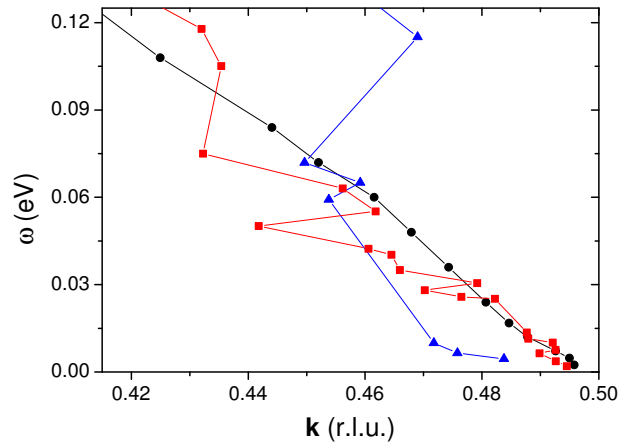}}}
\caption{The HWHM of the susceptibility maxima in reciprocal lattice units in PLCCO from Ref.~\protect\cite{Fujita06} ($x=0.11$, $T_c=25.5$~K, $T=6$~K, red squares), Ref.~\protect\cite{Wilson} ($x=0.12$, $T_c=21$~K, $T=7$~K, blue triangles), and in the present calculations with the parameters given in the text and $\delta=0.001$, $\Delta=9.6$~meV (black circles). Above $\omega\approx 0.1$~eV results of Ref.~\protect\cite{Wilson} give the dispersion of resolved incommensurate peaks.} \label{Fig1}
\end{figure}
The dispersion of the susceptibility $\chi''({\bf k}\omega)$ at half maximum, obtained in the present calculations, is shown in Fig.~\ref{Fig1}. This quantity can be compared with results of Refs.~\cite{Fujita11,Wilson,Fujita06}, in which an unresolved commensurate peak is observed in momentum cuts up to $\omega\approx 0.07$~eV \cite{Wilson} or even up to 0.17~eV \cite{Fujita11,Fujita06}. Therefore, in these works frequency dependencies of the half-width at half-maximum (HWHM) of these peaks are given. These dependencies are also shown in Fig.~\ref{Fig1}. As can be seen from this comparison, the calculations reproduce the experimentally observed cone-shaped dispersion with the apex point at $\omega=0$ and {\bf k=Q}. Moreover, the obtained $q$ widths of peaks are in reasonable agreement with those observed in Refs.~\cite{Fujita11,Fujita06}. This agreement is not a result of a $\Gamma$ fitting -- the $q$ widths depend only weakly on this parameter. As observed in Ref.~\cite{Wilson}, above $\omega\approx 0.1$~eV the commensurate maximum splits into incommensurate peaks, which disperse like spin waves.

\begin{figure}
\centerline{\resizebox{1.\columnwidth}{!}{\includegraphics{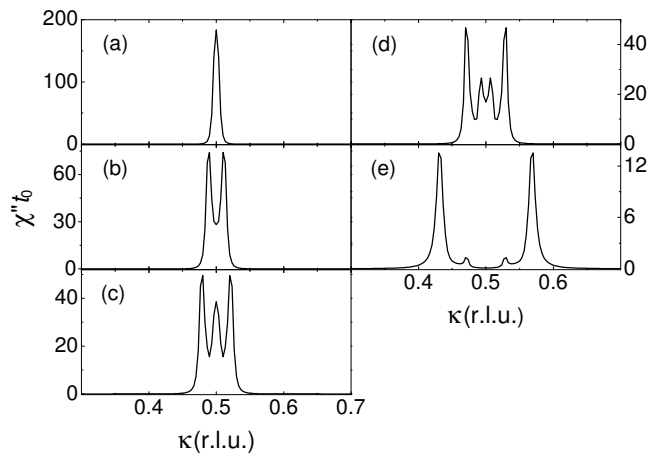}}}
\caption{The momentum dependence of $\chi''({\bf k}\omega)$ for $\omega=4.8$~meV (a), 16.8~meV (b), 36~meV (c), 48~meV (d) and 108~meV (e). The wave vector varies along a diagonal of the Brillouin zone, ${\bf k}=(\kappa,\kappa)$. Parameters are the same as in Fig.~\protect\ref{Fig1}.} \label{Fig2}
\end{figure}
\begin{figure}
\centerline{\resizebox{0.65\columnwidth}{!}{\includegraphics{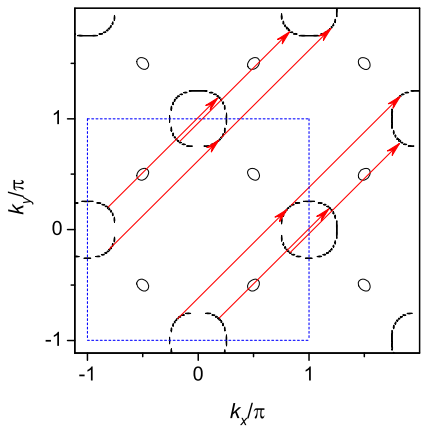}}}
\caption{The equi-energy contours of electron dispersion (\protect\ref{bands}), (\protect\ref{tbd}) for $\omega=\pm 6$~meV and $\Delta=3.6$~meV (black ovals). The blue dashed box is the boundary of the Brillouin zone. Red arrows show transitions making the main contributions to the polarization operator (\protect\ref{pi3}) for $\delta=0.001$, $\omega=1.2$~meV and ${\bf k=Q}$.} \label{Fig3}
\end{figure}

The calculated momentum cuts of $\chi''({\bf k}\omega)$ along a diagonal of the Brillouin zone are shown in Fig.~\ref{Fig2}. For small frequencies the susceptibility peaks at {\bf Q} [Fig. \ref{Fig2}(a)] -- the magnetic response is commensurate. This result is retained for other used values of $\Delta$ and $\delta$. As follows from the above formulas, the main reason for such a response is the electron band folding across the AF Brillouin zone border. This folding produces nested low-frequency equi-energy contours with the nesting vector {\bf Q} (see Fig.~\ref{Fig3}). Transitions between electron states on these contours make the main contribution to polarization operator (\ref{pi3}). As a consequence $-{\rm Im}\Pi({\bf k}\omega)$ in the numerator of the formula for $\chi''({\bf k}\omega)$ peaks sharply at {\bf Q}. With a weaker momentum dependence of the denominator at small $\omega$, the numerator controls the behavior of the susceptibility. The denominator also contributes to the appearance of the commensurate response, reaching the minimum at {\bf Q}. Notice that in $p$-type cuprates the situation is different -- in this case $-{\rm Im}\Pi({\bf k}\omega)$ has sharp peaks at incommensurate momenta. It is these peaks that produce the down-directed branch of the hourglass dispersion \cite{Sherman12}.

For the used parameters as the frequency exceeds 10~meV, the commensurate maximum splits into incommensurate peaks [Figs.~\ref{Fig2}(b)-(e)]. In the range $15\,{\rm meV}\lesssim\omega\lesssim 70$~meV peak intensities on the diagonals of the Brillouin zone exceed somewhat their value on the zone edge. For larger frequencies the susceptibility becomes nearly isotropic around {\bf Q}. As seen from Fig.~\ref{Fig2}, for moderate frequencies $\chi''({\bf k}\omega)$ consists of several closely spaced peaks. Apparently, these peaks merge into one broad maximum due to crystal inhomogeneity, as observed in experiments of Refs.~\cite{Fujita11,Wilson,Fujita06}.

\begin{figure}
\centerline{\resizebox{0.95\columnwidth}{!}{\includegraphics{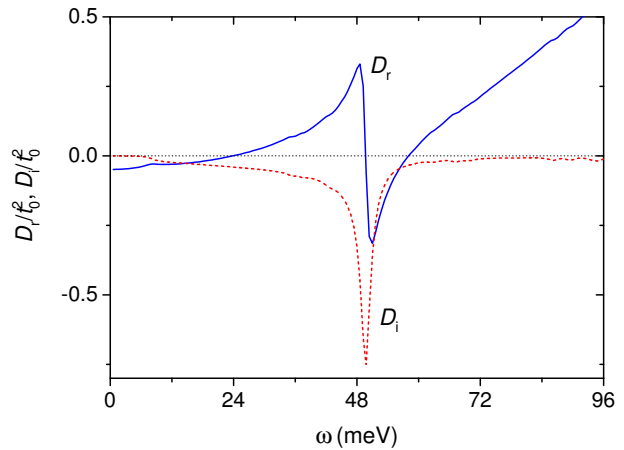}}}
\caption{The frequency dependencies of the real $D_r$ (the blue solid line) and imaginary $D_i$ (the red dashed line) parts of the denominator in Eq.~(\protect\ref{chi}) for ${\bf k}=(0.97\pi,0.97\pi)$, $\delta=0.001$ and $\Delta=4.8$~meV.} \label{Fig4}
\end{figure}
The peaks in Fig.~\ref{Fig2}(b)-(e) correspond to spin excitations. Susceptibility (\ref{chi}) coincides with the spin Green's function, and the vanishing real part of its denominator
\begin{equation}\label{pole}
D_r({\bf k}\omega)=\omega^2-\omega{\rm Re}\Pi({\bf k}\omega)-\omega^2_{\bf k}=0
\end{equation}
at a small imaginary part $D_i({\bf k}\omega)=\omega{\rm Im}\Pi({\bf k}\omega)$ defines the dispersion $\Omega_{\bf k}$ of these excitations. Figure~\ref{Fig4} demonstrates a graphical solution of Eq.~(\ref{pole}) for a wave vector near {\bf Q}. The real part has three zeros one of which falls into the region of large spin-excitation damping. Therefore, it is not seen in $\chi''({\bf k}\omega)$. Two other zeros correspond to maxima in Fig.~\ref{Fig2}(c)-(e). The zero with a lower frequency is similar to the respective zero of the real part of the susceptibility denominator in the hole-doped case \cite{Sherman12} -- with some correction due to ${\rm Re} \Pi({\bf k}\omega)$ it is close to $\omega_{\bf k}$. The zero at $\omega\approx 58$~meV arises due to the region of anomalous dispersion in $D_r({\bf k}\omega)$, which is related to the sharp minimum in $D_i({\bf k}\omega)$. Both peculiarities are consequences of the folded electron dispersion in this class of crystals. Small equi-energy contours shown in Fig.~\ref{Fig3} and dominant contribution of states on these contours into
$\Pi({\bf k}\omega)$ are inherent in these crystals for moderate frequencies. This leads to sharp minima in $D_i({\bf k}\omega)$ when the wave vector {\bf k} connects two such contours for a selected frequency.

\begin{figure}
\centerline{\resizebox{0.95\columnwidth}{!}{\includegraphics{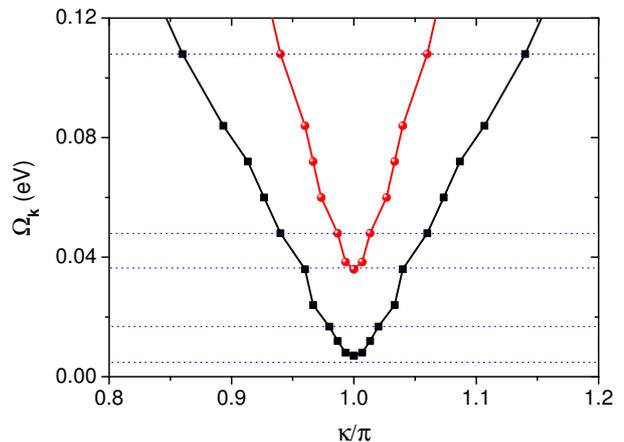}}}
\caption{The usual (black squares) and supplementary (red circles) dispersions of spin excitations. Parameters are the same as for Fig.~\protect\ref{Fig1}. Horizontal dotted lines indicate frequencies in panels of Fig.~\protect\ref{Fig2}. The wave vector varies along the zone diagonal, ${\bf k}=(\kappa,\kappa)$.} \label{Fig5}
\end{figure}
Figure~\ref{Fig5} demonstrates dispersions of two zeros of Eq.~(\ref{pole}), which fall into regions of small spin-excitation damping in Fig.~\ref{Fig4}. The locations of these zeros are close to the positions of maxima in Fig.~\ref{Fig2}(c)-(e). As mentioned above, the lower curve is close to the dispersion $\omega=\omega_{\bf k}$, and in this respect, it is similar to the spin-excitation dispersion in $p$-type cuprates. In the latter crystals, this dispersion forms the up-directed branch of the hourglass dispersion of susceptibility maxima. The difference between these compounds and the considered crystal is in the lowest frequency of the spin-excitation dispersion $\omega_r$. For the parameters of Fig.~\ref{Fig5} it is approximately equal to 6~meV, an order of magnitude smaller than in moderately doped $p$-type cuprates. Besides the usual spin-excitation branch, in the considered crystal there is a supplementary dispersion, which is connected with the second zero at higher frequencies. As mentioned above, its appearance is related to folded electron bands in $n$-type cuprates.

Thus, there are some similarity in the dispersions of susceptibility maxima in the two groups of cuprates -- one of the two up-directed branches in $n$-type crystals is similar to the upper hourglass branch in the hole-doped compounds. In the electron-doped cuprates the maxima of this branch are usually more intensive than peaks of the supplementary branch (see Fig.~\ref{Fig2}). This latter branch is the first difference between the two groups. The second difference is the behavior of $\chi''({\bf k}\omega)$ for $\omega<\omega_r$, where Eq.~(\ref{pole}) has no solutions and, therefore, there are no spin excitations [Fig.~\ref{Fig2}(a), the lowest horizontal line in Fig.~\ref{Fig5}]. As mentioned above, in this frequency range maxima in the momentum dependence of the susceptibility are caused by peaks in $-{\rm Im}\Pi({\bf k}\omega)$ in the numerator of the expression for $\chi''({\bf k}\omega)$. In $p$-type cuprates these peaks are at incommensurate momenta, which accounts for the down-directed branch of the hourglass dispersion \cite{Sherman12}. In this branch, with decreasing $\omega$ maxima of $\chi''({\bf k}\omega)$ move away from {\bf Q}. In $n$-type crystals $-{\rm Im}\Pi({\bf k}\omega)$ peaks at the AF momentum due to electron bands folded across the AF Brillouin zone border. As a consequence in this case the susceptibility dispersion has the shape of a cone with the apex point at $\omega=0$, ${\bf k=Q}$ (see Fig.~\ref{Fig1}).

\begin{figure}
\centerline{\resizebox{0.75\columnwidth}{!}{\includegraphics{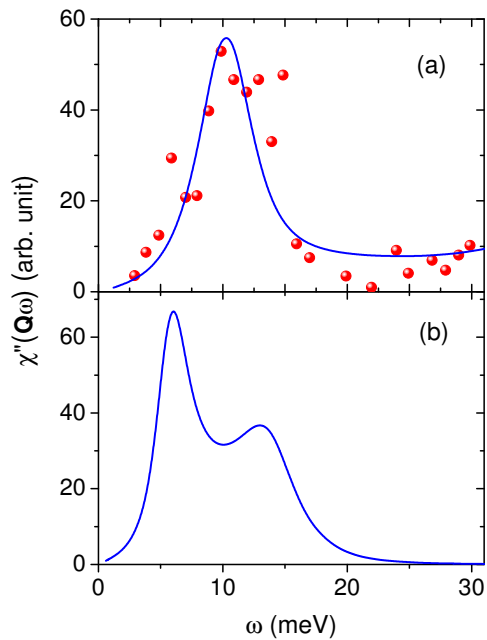}}}
\caption{The frequency dependence of the susceptibility at {\bf k=Q}. (a) Experimental data in PLCCO with $x=0.12$ and $T_c=24$~K at $T=2$~K \protect\cite{Zhao} (circles) and the calculated $\chi''$ for $\delta=4\times 10^{-3}$ and $\Delta=9.6$~meV (the curve). (b) the calculated susceptibility for $\delta=5\times 10^{-4}$ and $\Delta=4.8$~meV.} \label{Fig6}
\end{figure}
The frequency dependence of the susceptibility at the AF wave vector is shown in Fig.~\ref{Fig6}. In panel (a) the calculated susceptibility is compared with experimental results from Ref.~\cite{Zhao}. In these calculations, the parameter $\delta$ was chosen to fit the location of the maximum to the experimental one. In the calculated susceptibility, the maximum approximately coincides with the bottom of the usual spin-excitation branch in Fig.~\ref{Fig5}. The bottom of the supplementary dispersion is seen as a weaker maximum at $\omega\approx 45$~meV. Although the measurements \cite{Zhao} were limited to $\omega\lesssim 30$~meV, some growth in the experimental susceptibility may be also observed near the upper edge of this frequency window. In the local magnetic susceptibility measured in Ref.~\cite{Fujita06} a shoulder or a maximum may be revealed at $\omega\approx 30-45$~meV, which can be also related to the supplementary spin excitations. In recent work \cite{Zhao11} a PLCCO crystal with $x=0.12$ and $T_c=21$~K was investigated. In comparison with the $T_c=24$~K sample of Ref.~\cite{Zhao} the $T_c=21$~K sample has smaller superconducting gap $\Delta\approx 5.5$~meV and is presumably located much closer to the boundary of the long-range AF order. Therefore, this crystal has to be characterized by a much smaller parameter $\delta$ than the $T_c=24$~K sample. Two maxima at 2 and 9.5~meV were observed in the local susceptibility at $T=2$~K. Calculations in Fig.~\ref{Fig6}(b) carried out with smaller than in panel (a) values of the parameters $\Delta$ and $\delta$ reproduce qualitatively this experimental spectrum. Maxima in this figure correspond to the two branches of the spin-excitation spectrum. Notice that in Ref.~\cite{Zhao11} a maximum at $\omega=2$~meV was also observed in the $T_c=24$~K sample. However, this maximum was much weaker than the peak at 10.5~meV and the maximum at 2~meV in the $T_c=21$~K sample. I could not obtain such kind of spectrum. It is conceivable that the weak maximum is connected with sample inhomogeneity.

\section{Conclusion}
In this work, the two-dimensional $t$-$J$ model and the Mori projection operator technique were used for interpreting some peculiarities of the magnetic susceptibility in the optimally doped PLCCO crystal. The electron dispersion derived from photoemission data was applied in the calculations. The dispersion is characterized by the band folding across the antiferromagnetic Brillouin zone border. This band folding plays a central role in the formation of the low-frequency commensurate response and in the appearance of the supplementary branch of spin excitations. The origin of these peculiarities is related to small equi-energy contours of the electron dispersion, which are the consequence of the band folding. For small frequencies, the nesting vector of the contours is the antiferromagnetic momentum ${\bf Q}=(\pi,\pi)$. Therefore, for such frequencies the imaginary part of the polarization operator $-{\rm Im}\Pi({\bf k}\omega)$ peaks sharply at {\bf Q}. This operator is in the numerator of the formula for the susceptibility $\chi''({\bf k}\omega)$, and at a weaker momentum dependence of the denominator, it controls the behavior of the susceptibility, leading to the commensurate response. For moderate frequencies, the peak in $-{\rm Im}\Pi({\bf k}\omega)$ splits and shifts to incommensurate wave vectors. An anomalous dispersion in the real part of the polarization operator, which is connected with the peak, leads to the appearance of three poles in the spin Green's function. Two of these poles fall into regions of a small spin-excitation damping and correspond to the usual and supplementary spin-excitation branches. The latter branch nested into the former leads to a comb of closely spaced peaks in momentum cuts. Presumably this structure is not resolved in experiments, being seen as a broad commensurate peak up to $\sim 100$~meV. It is very likely that the supplementary branch was already observed in the recent measurements of the frequency dependence of the susceptibility.

Comparing with moderately doped $p$-type cuprates a common element in dispersions of susceptibility maxima can be revealed -- it is branches of usual spin excitations. However, in the PLCCO there exists also the supplementary branch nested into the usual one. Besides, in the PLCCO the bottom frequency of the usual branch $\omega_r$ is an order of magnitude smaller than in hole-doped crystals. For $\omega<\omega_r$ the behavior of the susceptibility in the two groups of compounds is drastically different. In both groups this behavior is governed by $-{\rm Im}\Pi({\bf k}\omega)$ in the numerator of the susceptibility formula. However, if in the PLCCO this quantity peaks at {\bf Q}, which leads to the commensurate low-frequency response, in hole-doped crystals $-{\rm Im}\Pi({\bf k}\omega)$ reaches a maximum value at incommensurate momenta, and the low-frequency response is incommensurate.

\begin{acknowledgement}
This work was supported by the European Regional Development Fund (Centre of Excellence "Mesosystems: Theory and applications", TK114) and by the Estonian Scientific Foundation (grant ETF9371).
\end{acknowledgement}

\end{document}